\documentclass[twocolumn]{emulateapj}
\usepackage[latin2]{inputenc}

\mathchardef\mhyphen="2D

\newcommand{\bssC}{\textbf{\textsf{C}}}
\newcommand{\R}{\mathcal{R}}
\newcommand{\avg}[1]{\left\langle{#1}\right\rangle}

\newcommand{\hmpc}{\,$h^{-1}$\,Mpc}
\newcommand{\hgpcnosp}{$h^{-1}$\,Gpc}
\newcommand{\hmpcnosp}{$h^{-1}$\,Mpc}
\newcommand{\ihmpc}{\,$h$\,Mpc$^{-1}$}

\newcommand{\Var}{{\rm Var}}

\newcommand{\sns}{(S/N)$^2$}

\newcommand{\varc}{\sigma^2_{\rm cell}}
\newcommand{\varcs}{\sigma^4_{\rm cell}}

\newcommand{\cngij}{C^{\rm NG}_{ij}}
\newcommand{\sqninj}{(N_iN_j)^{1/2}}
\chardef\til=`\~

\begin{document}

\title{Removable Matter-Power-Spectrum Covariance from Bias Fluctuations}

\author{Mark C. Neyrinck\\
{\rm \small 
Department of Physics and Astronomy, The Johns Hopkins University, Baltimore, MD 21218, USA}
}

\begin{abstract}
We find a simple, accurate model for the covariance matrix of the
real-space cosmological matter power spectrum on slightly nonlinear
scales ($k\sim$ 0.1-0.8\ihmpc\ at $z=0$), where off-diagonal matrix
elements become substantial.  The model includes a multiplicative,
scale-independent modulation of the power spectrum.  It has only one
parameter, the variance (among realizations) of the variance of the
nonlinear density field in cells, with little dependence on the cell
size between 2-8\hmpc.  Furthermore, we find that this extra
covariance can be modeled out by instead measuring the power spectrum
of $\delta/\sigma_{\rm cell}$, i.e.\ the ratio of the overdensity to
its dispersion in cells a few Mpc in size.  Dividing $\delta$ by
$\sigma_{\rm cell}$ essentially removes the non-Gaussian part of the
covariance matrix, nearly diagonalizing it.
\end{abstract}


\section{Introduction}
The power spectrum of density fluctuations (and its
Fourier dual, the correlation function) is the most common statistic
used to characterize the large-scale structure of the Universe, and
contains much information about cosmological parameters.  In linear
theory, the power spectrum's covariance matrix is diagonal, a practical
advantage over the correlation function.

However, in the `translinear' regime, the first decade or so in
wavenumber smaller than fully linear scales, nonlinearity grows in the
density field.  The dispersion $\sigma_{\rm cell}$ in the nonlinear
overdensity field $\delta$ in cells of translinear size $2\pi/k\sim
4$-60\ihmpc\ increases from about 1 to 10 to 20, as $k$ decreases from
60 to 8 to 4\ihmpc, as measured in the simulations analyzed below.
(Note that the dispersion in the {\it linearly} evolved density field
in 8-\hmpcnosp\ cells, $\sigma_8\approx 0.8$.)

The growing nonlinearity in $\delta$ incites a substantial
non-Gaussian component in its power-spectrum covariance matrix
\citep[][T09]{mw,szh,t09}.  The non-Gaussian covariance leads to a
translinear plateau in Fisher information
\citep{rh05,rh06,nsr06,ns07}.  When using the power spectrum of
$\delta$ to constrain a parameter such as the initial power spectrum
amplitude, modes in the translinear regime are highly correlated,
giving little additional constraining power when analyzed with
larger-scale modes.

A few methods have been proposed that mitigate this problem, to
varying degrees: removing large halos from a survey
\citep{ns07}; applying a logarithmic or
rank-order-Gaussianizing transform to the density field before
measuring the power spectrum \citep{nss09,seo,yuyu}; and
nonlinear wavelet Wiener filtering \citep{zhang11}.

The non-diagonality of the power-spectrum covariance matrix is not
just a problem for parameter estimation.  Without a theoretical model,
off-diagonal elements must be estimated numerically, often
cumbersomely.  Even a covariance matrix estimated with 5000
simulations leaves residual noise (T09; see Fig.\ \ref{fig:tak}
below).  Also, the results often depend on the method used to measure
the covariance.  For example, spatial jackknife resampling
systematically underpredicts the covariance \citep{nbgc09}.  Another
strategy, which we employ, applies a large set of sinusoidal weighting
functions to a single density field \citep[][HRS]{hrs}.  The HRS
weightings are designed to be optimally friendly to Fourier modes
smaller than the sinusoids themselves, unlike windows with sharp
configuration-space edges.  Any window function introduces `beat
coupling' between small and large-scale modes; this contributes
off-diagonal covariance beyond what is found in an ensemble of
unweighted, periodic simulations.  For the sinusoidal weightings, the
beat coupling likely does not exceed that for more-disruptive, sharp
window functions, often present in observations.  Another recent
technique to generate low-noise covariance matrices from a relatively
small number of simulations makes use of `mode-resampling'
\citep{scfs11}, taking advantage of the plethora of small-scale modes
available.

`Shrinkage estimation' \citep{ps} is possible if an approximate model
is known for the covariance; this gives an optimal combination of a
noiseless but idealized model, and a complete but noisy measured
covariance matrix.  One popular theoretical model for the covariance
uses a halo-model (HM) framework \citep{coorayhu}.  The full HM
covariance includes up to four-halo terms, and requires knowledge of
the halo-halo power spectrum, bi- and trispectrum.  Unfortunately,
even the halo-halo power spectrum currently lacks a satisfying
analytical model \citep{sss07}.  Leading-order perturbation-theory
power, bi-, and trispectra can be used, but on translinear scales they
are likely inaccurate for halo-halo correlations, making their
nontrivial implementation possibly not worth the trouble.  Thus, in
practice often only the simpler 1-halo term is used.

In this paper, we explore a simple, accurate model for translinear
covariance from a fluctuating multiplicative bias.  \citet{sato09}
considered a term with a rather similar form to this, called `halo
sampling variance' (HSV), consisting of a variance in bias multiplying
one-halo power-spectrum terms.  They found that including this HSV,
which, unlike us, they ascribe to modes outside the survey volume, was
necessary to model the covariance of the weak-lensing convergence
power spectrum accurately in a halo-model approach.

\section{Results}

We measured real-space power spectra and covariance matrices from the
Coyote Universe \citep{coyote1,coyote2,coyote3} suite of 37
simulations.  Each of these has a different set of cosmological
parameters, each of them a plausible (given current observational
constraints) concordance $\Lambda$CDM model.  The simulations occupy
an orthogonal-array-Latin-hypercube in the five-dimensional parameter
space $\Omega_mh^2\in[0.12,0.155]$, $\Omega_bh^2\in[0.0215,0.0235]$,
$n_s\in[0.85,1.05]$, $w\in[-1.3,-0.7]$, $\sigma_8\in[0.61,0.9]$.  The
remaining cosmological parameters, e.g.\ $h$, are set to match the
tight cosmic microwave background (CMB) constraint on the ratio of the
last-scattering-surface distance to the sound-horizon scale.  The
1024$^3$-particle simulations have box size 1300 Mpc, fixed in Mpc
(not \hmpc) to roughly line up baryon-acoustic-oscillation (BAO)
features in $k$ among different cosmologies.  Their resolution is
sufficient for power-spectrum measurements accurate at sub-percent
level at scales down to $k=1$\ihmpc.  We measured power spectra on
grids with nearest-grid-point density assignment, not correcting for
the pixel window function on the $128^3$-$512^3$ grids we use.

The variable set of cosmologies prevented estimating covariances by
directly comparing power spectra among simulations.  So, we measured
covariances within each simulation separately, using 248 (combinations
of no weighting, and the first and second overtones in each Cartesian
direction) sinusoidal weightings prescribed by HRS.  We then averaged
the covariance matrices from different cosmologies together to reduce
noise.  Arguably, mixing cosmologies in this way is a good thing,
decreasing the dependence on cosmological parameters.

We did not use the covariance of the power spectrum $P$ itself, but of
$\ln P$, defining $C_{ij}\equiv\avg{\Delta\ln P_i\Delta\ln
  P_j}=\avg{\Delta P_i\Delta P_j}/\avg{P_i}\avg{P_j}$.  Averaging this
covariance matrix across cosmologies is more stable than the
covariance matrix of $P$ itself.  We also define the quantity
$T_{ij}\equiv C_{ij}\sqninj - 2\delta^K_{ij}$, where $\delta^K_{ij}$
is a unit matrix, and $N_i$ is the number of modes in bin $i$.
$T_{ij}$, zero for a Gaussian random field, is a measure of the
non-Gaussian component of a covariance matrix, normalized to the
Gaussian component.  Conveniently for plotting purposes, it has
roughly uniform noise properties.

\begin{figure}
  \begin{center}
    \includegraphics[scale=0.4]{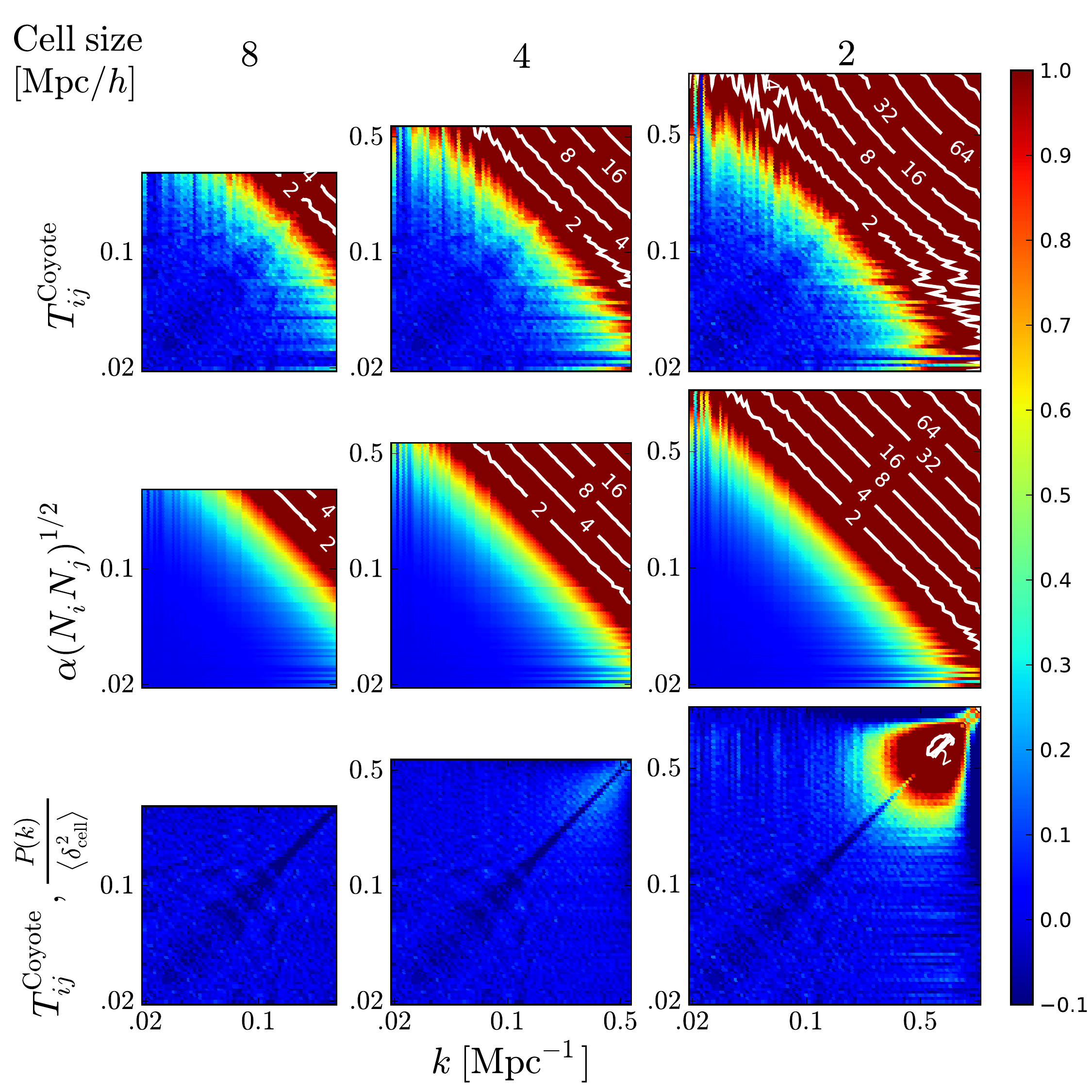}
  \end{center}  
  \caption{ Pixels are colored according to the quantities listed on
    the left edge of the plots; for quantities exceeding 1, contours
    appear at successive powers of 2. {\it Top row}.  The normalized
    non-Gaussian part of the covariance matrix $T_{ij}$. {\it Middle
      row}.  An approximation to $T_{ij}$, using
    $\alpha=\Var(\varc)/\varcs$, measured from each simulation's
    ensemble of sinusoidal weightings, and geometrically averaged
    across simulations.  {\it Bottom row.} $T_{ij}$ for the power
    spectrum normalized by the cell-density variance,
    $P(k)/\sigma^2_{\rm cell}$, equivalent to the power spectrum of
    $(\delta/\sigma_{\rm cell})$.  (The displayed range of $k$ includes
    modes not directly modulated by the weightings, and below the
    Nyquist $k$.  The cell sizes are 1300/128, 1300/256 and 1300/512
    Mpc, roughly the listed values in \hmpcnosp.)  }
  \label{fig:pv_corr_res}
\end{figure}

The top row of Fig.\ \ref{fig:pv_corr_res} shows $T_{ij}$, measured
from the simulations.  As the middle row shows, $T_{ij}$ is
well-approximated by $T_{ij}=\alpha \sqninj$, for
$\alpha=\Var(\varc)/\varcs$.  Here $\Var(\varc)$ is the variance of
the measured (not linear-theory) cell-density variance $\varc$ among
sinusoidally weighted density fields.  $\alpha$, proportional to a
variance, is inversely proportional to the volume $V=(1.3\ {\rm
  Gpc})^3/2$ (the sinusoidal weightings effectively halve the volume).

The approximate form of this covariance matrix is that of a Gaussian
field, with additional power proportional to a scale-independent bias.
Consider a field with power spectrum $P_i=b\avg{P_i}+P^G_i$.  Here
$P^G$ is the power spectrum of a Gaussian random field such that
$\avg{P_i}=\avg{P^G_i}$, and $b$ is a variable scale-independent bias,
uncorrelated with fluctuations in $P^G$, with $\avg{b}=0$.  (Suppose
$b> -1$, e.g.\ is lognormally distributed, to avoid $P_i<0$.)  $P$'s
covariance matrix is
\begin{equation}
  C_{ij}=\avg{b^2}\avg{P_i}\avg{P_j} + \avg{\Delta P^G_i\Delta P^G_j}.
\end{equation}
$\avg{\Delta P^G_i\Delta P^G_j}=2\delta^K_{ij}P_i^2/N_i$, so
$T_{ij}=\avg{b^2}\sqninj$.  Neglecting fluctuations from $P^G$,
$\varc=\avg{b^2}\varcs$.  Here the proportionality constant
$\alpha=\avg{b^2}$.

With the insight that the covariance is from bias fluctuations,
perhaps the covariance can be removed. Indeed, in the bottom row of
Fig.\ \ref{fig:pv_corr_res}, we show that the translinear covariance
largely vanishes if the power spectrum from each density field is
divided by $\varc$, measured from that field.  This is equivalent to
measuring the power spectrum of ($\delta/\sigma_{\rm cell}$).  The
reduction in covariance is dramatic for cell sizes $\gtrsim4$\hmpc.
Using 2-\hmpcnosp\ cells, there is significant residual covariance on
small scales, but the covariance is still much smaller than in the top
rows.  Interestingly, $\alpha$ is quite consistent over this factor of
4 in cell size; with 2, 4, and 8\hmpc\ cells, $\alpha\times10^4$=1.29,
1.37 and 1.22.

\begin{figure}
  \begin{center}
    \includegraphics[scale=0.4]{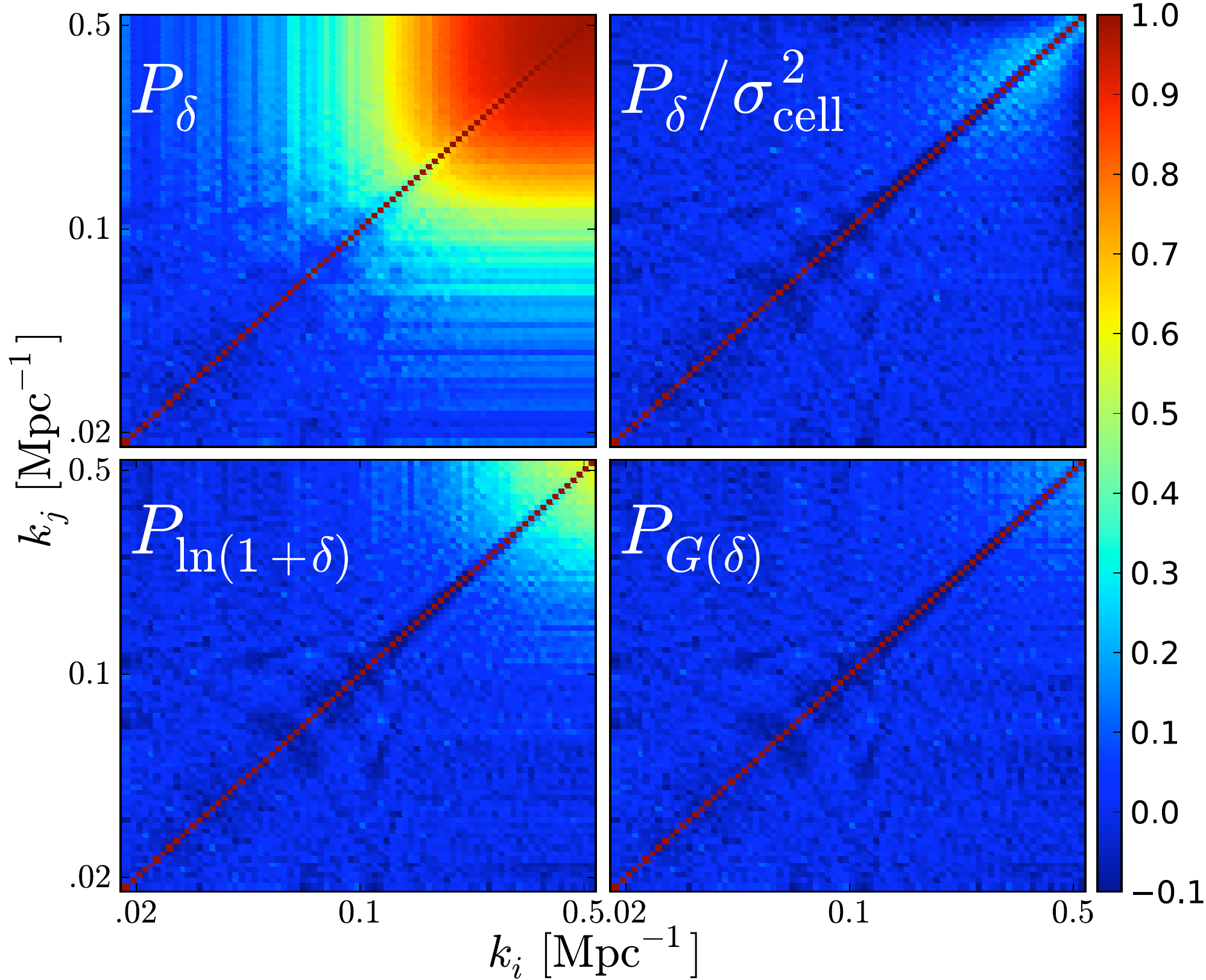}
  \end{center}  
  \caption{Correlation matrices of various power spectra measured from
    the Coyote Universe simulations.  $P_\delta$ is the conventional
    power spectrum; $P_\delta/\varc$ is the power spectrum,
    dividing by the variance of cell densities; $P_{\ln(1+\delta)}$ is
    the power spectrum of the log-density; $P_{G(\delta)}$ is the
    power spectrum of the Gaussianized density.  }
  \label{fig:corrmatrix}
\end{figure}

Fig.\ \ref{fig:corrmatrix} shows a commonly plotted measure of the
covariance, the correlation matrix $C_{ij}/(C_{ii}C_{jj})^{1/2}$.
$P_\delta/\varc$ has smaller covariance even than the $\ln(1+\delta)$
power spectrum (NSS09), surpassed in diagonality only by the
Gaussianized power spectrum (NSS09).  Hereafter, all of the Coyote
Universe results use a 256$^3$ grid, with a cell size of $\sim
4$\hmpc.

We note that the form of the $\ln(1+\delta)$ covariance matrix is
similar to the $\delta$ covariance matrix, except that the constant
$\alpha$ is given by fluctuations in the variance of $\ln(1+\delta)$
in cells.  This sheds light on the reduction in covariance from
transforming the density to give a more-Gaussian 1-point distribution.
By definition, Gaussianization reduces higher moments, in particular
the kurtosis, and thus the variance in the variance.  In our model,
then, the full power-spectrum covariance is reduced.

\subsection{Comparison with the halo model}
\begin{figure}
  \begin{center}
    \includegraphics[scale=0.29]{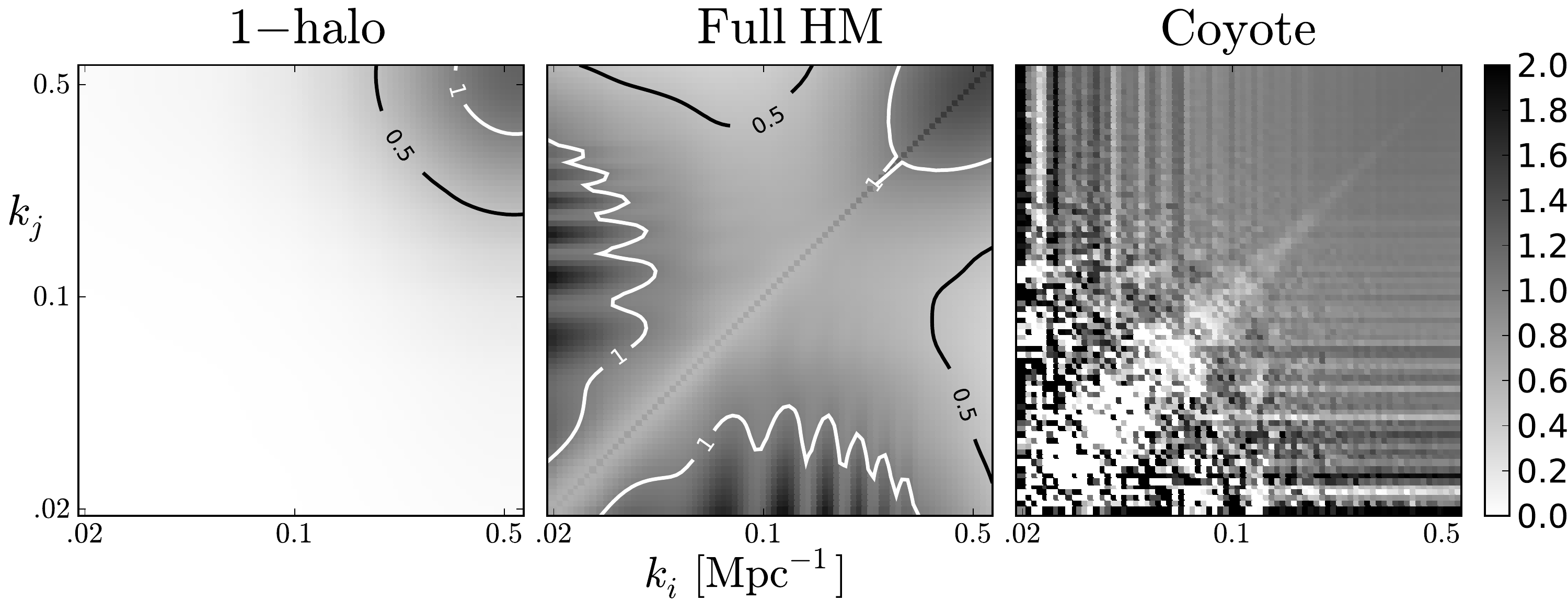}
  \end{center}  
  \caption{The non-Gaussian part of the covariance, $\cngij$, in the
    halo model, including just the 1-halo term (left panel), and with
    all terms (middle panel). $\cngij$ as measured in the Coyote
    Universe simulations appears in the right panel.  In all cases,
    $\cngij$ is divided by $\alpha=\Var(\varc)/\varcs$, measured from
    the simulations.  }
  \label{fig:halocov}
\end{figure}

Fig.\ \ref{fig:halocov} shows the quantity $\cngij\equiv
C_{ij}-2\delta^K_{ij}/\sqninj$, the covariance minus the Gaussian
component; this should be a constant in our $\alpha$ model.  We
compare $\cngij$ as measured from the Coyote Universe to predictions
in the HM, both using all terms, and only the 1-halo term.  For the HM
covariance matrix we use our {\scshape CosmoPy} implementation
\citep{nsr06,ns07,ns08}.  In all panels, $\cngij$ is
normalized by dividing by $\alpha$ as measured in the simulations.

Indeed, at large $k$, the measured $\cngij/\alpha\approx 1$,
conforming to the form above.  For the HM panels, we set
$N_i\propto k^2$ exactly; i.e.\ fluctuations in $N_i$ from the finite
lattice are suppressed for clarity.  The detailed features in the
HM $\cngij$ are different than that measured.  On large
scales, where there is much noise in the measured $\cngij$, there is
barely a hint of the perturbation-theory trispectrum wiggles from BAO \citep{ns08}.  This is not surprising; in that paper it
took a few hundred Gpc-scale simulations to get a clear signal.

While our simple model involving $\alpha$ seems more accurate than the
HM covariances on translinear scales, the HM does not fail utterly;
the full-HM covariance is typically within a factor of two of that
measured.  Given that we did not adjust HM parameters or polyspectrum
assumptions to optimize the fit, its agreement is not bad.  The
agreement with the commonly used 1-halo covariance, however, is poorer
outside the large-$k$ corner.  This vanilla 1-halo model assumes a
halo mass function with uncorrelated Poisson fluctuations in each mass
bin \citep{nsr06}; perhaps including mass-function covariance would improve
agreement.

\subsection{Comparison with other results}
\begin{figure}
  \begin{center}
    \includegraphics[scale=0.44]{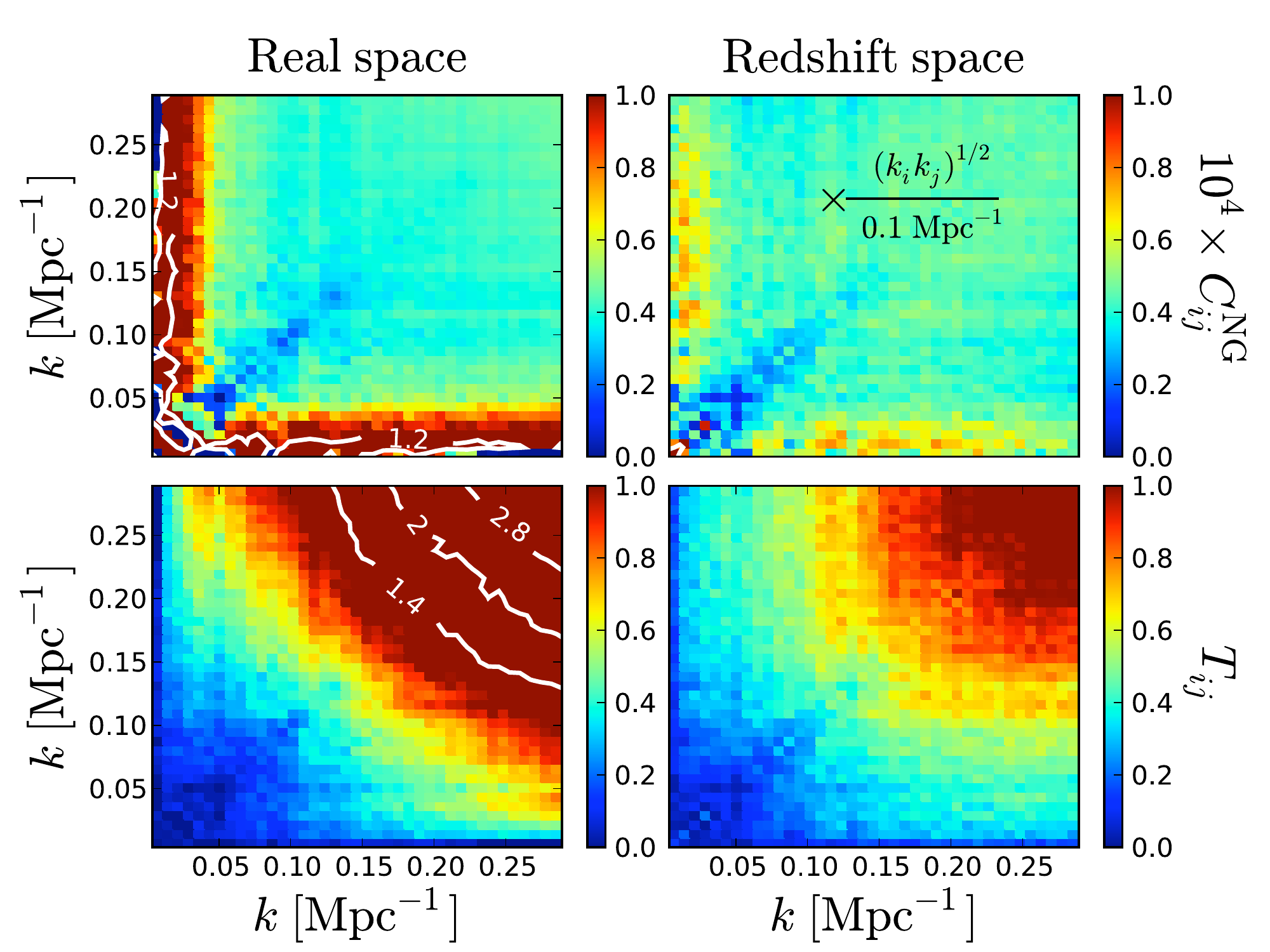}
  \end{center}  
  \caption{Real- and redshift-space non-Gaussian covariances
    $\cngij$ and $T_{ij}$ measured from an ensemble of 5000
    simulations by T09.  The redshift-space $\cngij$ is additionally
    multiplied by an empirically found factor $\propto(k_ik_j)^{1/2}$
    to become roughly constant for large $k$.  Note the linear $k$ scale.}
  \label{fig:tak}
\end{figure}

It is well worth checking our result with other covariance
measurements, given our averaging procedure over cosmologies, and the
`beat-coupling' induced by our sinusoidal weightings.
Fig.\ \ref{fig:tak} shows measurements by T09 of $T_{ij}$ and $\cngij$
from their suite of 5000 $\Lambda$CDM simulations of volume 1
(\hgpcnosp)$^3$, each with 256$^3$ particles.

For $k\gtrsim 0.04$\ihmpc, the real-space $\cngij$ is roughly
constant, in accordance with our model.  However, for smaller $k$,
$\cngij$ roughly doubles from its smaller-scale value; unsurprisingly,
other effects take over on the largest scales.  As in the Coyote
Universe results, at BAO scales there are interesting depressions in
the covariance near the diagonal, where the covariance even dips
slightly negative.  In Fig.\ \ref{fig:tak} there is also a suggestion
of periodic BAO bumps along the axes.

In redshift space, we found empirically that $\cngij$ instead falls
off roughly as $(k_ik_j)^{-1/2}$.  We do not speculate here on why it
has this form, but the reduced redshift-space covariance on small
scales, noted by T09, is not surprising given the more-Gaussian
redshift-space density 1-point distribution, smeared by redshift distortions
\citep{nss11}.

\subsection{Fisher information}

\begin{figure}
  \begin{center}
    \includegraphics[scale=0.31]{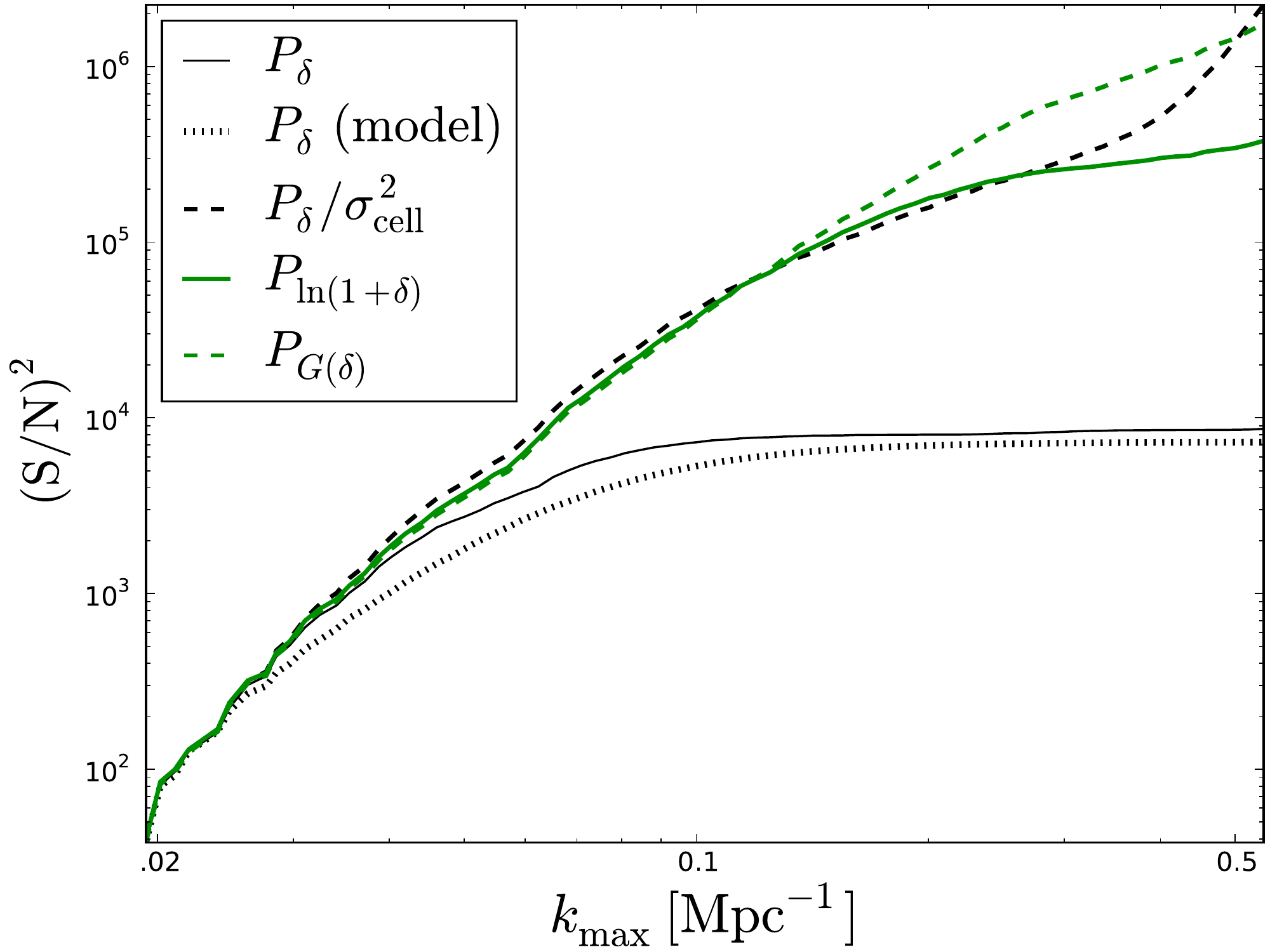}
  \end{center}  
  \caption{The intrinsic Fisher information, \sns, in various power
    spectra, measured from the Coyote Universe simulations.  The
    dotted curve uses our model for the $P_\delta$ covariance from
    fluctuations in a scale-independent multiplicative bias.
  }
  \label{fig:coyoteinfo}
\end{figure}

Fig.\ \ref{fig:coyoteinfo} shows signal-to-noise, \sns, curves for
the various power spectra.  This \sns\ can be thought of as the
effective number of independent modes over a range of wavenumber bins
$\R$.
\begin{equation}
  ({\rm S/N})^2(\R) = \sum_{i,j\in \R} (\bssC_\R^{-1})_{ij},
  \label{eqn:inforange}
\end{equation}
where $\bssC_\R$ is the covariance matrix over $\R$.  Here, the bins
in $\R$ vary from $k_{\rm min}$, the largest modes not directly
modulated by the sinusoidal weightings, to $k_{\rm max}$.

Compared to NSS09, the range of $k$ plotted here is shifted smaller by
a factor of 2 by the double box size, and there is reduced noise, but
the trends are the same as found there.  In fact, $P_\delta/\varc$ is
comparable to $P_{G(\delta)}$, the power spectrum of the Gaussianized
density.  The \sns\ in $P_\delta/\varc$ ramps up quickly at the
smallest scales, because the power spectra are effectively pinned
together there when they are divided by $\varc$, resulting in tiny
(co)variance.  We also show the \sns\ curve for our $\cngij=\alpha$
model, which approximates the $P_\delta$ curve well.

\sns\ is a good single measure of intrinsic Fisher information, but
without derivative terms of the power spectrum with respect to
cosmological parameters, the connection with parameter estimation is
vague.  For $P_\delta$, \sns\ is roughly the information in $\ln
\sigma_8^2$, unmarginalized over other parameters, since on linear
scales $\partial P_\delta(k)/(\partial \ln \sigma_8^2) =1$, and it
stays of order unity on non-linear scales, reaching $\sim 2$ \citep{ns07}.
However, at least for 8-\hmpcnosp\ cells, $P_\delta/\varc$ contains
little information about $\sigma_8^2$.  (If the measured power
spectrum were linear, obviously all sensitivity to $\sigma_8^2$ would
be lost in dividing by it.)  For $P_{G(\delta)}$, too, a free
parameter is the variance in the Gaussianized density.  In the present
measurements, as in \citet{nss11}, the $G(\delta)$ variance is fixed, pinning
down the power spectrum at small scales in a similar manner as with
$P_\delta/\varc$.  The situation with $P_{\ln(1+\delta)}$ is
in-between; on small scales it is allowed to fluctuate as it pleases
(which is to a much lesser degree than $P_\delta$), but on large
scales, $0<\partial P_{\ln(1+\delta)}/(\partial \ln \sigma_8^2)<1$
because of large-scale bias produced by the log transform.

An analysis of these issues \citep{nprep} is essential for these
modified power spectra to be used for cosmological constraints.  One
might think that the insight that $P_\delta$'s covariance is largely
from fluctuations in a multiplicative bias might imply that this
covariance is unimportant for parameters that depend on the
power-spectrum shape, such as the tilt, or the BAO scale.  However, as
shown in Fig.\ \ref{fig:tak}, the form we have found for the
translinear covariance does not really extend to fully linear scales,
leaving some room for shape fluctuations at the interface (occupied by
BAO at $z=0$) between the linear and translinear regimes.

\section{Conclusion}

We show that a fluctuating scale-independent multiplicative bias
provides quite an accurate model for the covariance matrix of the
real-space dark-matter power spectrum on translinear scales.  The
non-Gaussian part of the covariance
$\cngij\approx\alpha=\Var(\varc)/\varcs$, a constant measure of the
variance (among realizations) of the variance (within a realization) of
the measured nonlinear density field.  $\alpha$ is rather insensitive
to the cell size used to measure it, at least within a range of
2-8\hmpc.  $\alpha$ is not entirely trivial to measure, but it is much
easier than the full covariance.  For example, it could be estimated
using sufficiently large simple sub-volumes, with little worry about
edge effects.

Furthermore, at least in real space, this translinear covariance can
largely be removed, or modeled out, by dividing the power spectrum by
$\varc$.  This is equivalent to measuring the power spectrum of
$\delta/\sigma_{\rm cell}$.  However, this changes the sensitivity to
cosmology, for example giving up information on the power-spectrum
amplitude.

The simplicity of this model suggests that it may be applicable not
only to dark matter, but perhaps to galaxies, or even to CMB
anisotropies, where pushing to smaller scales dredges up some
non-Gaussian covariance in the power spectrum of Sunyaev-Zel'dovich
effect anisotropies \citep{shaw09}.  Shot noise would complicate the
situation for galaxies, but possibly this component could be
accurately approximated with a Poisson term and subtracted.

\acknowledgments We thank Istv\'an Szapudi, Alex Szalay, Martin White
and S\'ebastien Heinis for helpful discussions, Ryuichi Takahashi for
sharing the covariance matrices used in Fig.\ \ref{fig:tak}, and
Katrin Heitmann and Adrian Pope for help accessing the Coyote Universe
simulations.  We are grateful for support from the W.M.\ Keck and the
Gordon and Betty Moore Foundations.

\bibliographystyle{hapj}
\bibliography{refs}

\end{document}